\documentclass[prl,twocolumn,showpacs,preprintnumbers,amsmath,amssymb]{revtex4}
\usepackage{amsfonts}
\usepackage{graphicx}
\usepackage{dcolumn}
\usepackage{float}
\usepackage{bm}

\newcommand{\com}[1]{}

\begin{document}

\title{On the probability of quantum phase slips in superconducting nanowires}

\author{D. Mozyrsky}\email{mozyrsky@lanl.gov}

\affiliation{Theoretical Division (T-4), Los Alamos National Laboratory, Los
Alamos, NM 87545, USA}

\date{\today}

\begin{abstract}

The paper discusses mechanisms for decay of supercurrents in ultrathin superconducting wires driven by quantum fluctuations.
We argue that momentum conservation strongly suppresses probability of such decay and estimate the rates for two
decay channels: potential scattering of condensate due to the disorder and attenuation of the plasmon mode due to
the presence of normal component, i.e., Ohmic losses. We find that while both mechanisms yield non-zero decay rates,
their values are too small to provide any substantial contribution to the resistivity of the wires. The rate associated
with the latter mechanism, however, is much greater, and it is possible that under the appropriate conditions dissipation
may lead to appreciable enhancement of quantum phase slip transitions.

\end{abstract}

\pacs{03.75.Kk, 37.10.Gh, 85.25.Cp}

\maketitle

Properties of superconducting nanowires at ultralow temperatures have been a subject of extensive theoretical and experimental studies for several decades. It is believed that such properties are controlled by proliferation of the so-called quantum phase slips (QPS), - topological fluctuations resulting in discontinuous changes of phase of the superconducting order parameter driven by quantum fluctuations \cite{demler}. Such QPS are expected to lead to finite resistivity of superconducting wires at temperatures much lower than critical temperature ($T_c$) and, under certain conditions, to superconductor-insulator quantum phase transition \cite{demler, zaikin, prokof}.

Despite theoretical predictions that QPS should be observed in sufficiently thin nanowires \cite{zaikin}, their experimental observation turned out to be rather difficult and inconclusive. While initial experiments seem to have demonstrated the existence of QPS fluctuations in MoGe nanowires \cite{bez}, subsequent measurements did not confirm these observations \cite{exp}. More recent measurements, though, have suggested that QPS might have been observed in nanowires carrying sufficiently high bias current, close to the value of critical current \cite{beznat}.

In this paper we estimate the rate of the QPS in superconducting nanowires. We find that this rate is controlled by two factors: (1) semiclassical exponent due to the quantum tunneling of the order parameter through an effective energy barrier, which arises due to the suppression of the superconducting density at the QPS cores; (2) a probability for the superflow to change its momentum by $h\rho/2$, where $\rho$ is 1-dimensional (1D) density of conduction electrons. We argue that in experimentally accessible regimes the latter factor is responsible for major suppression of the QPS rates. Particularly we estimate the rate for two mechanisms of momentum relaxation, such as scattering of the condensate at the disorder potential, e.g. Ref. \cite{khleb2, prokof}, and due to the dissipation of the plasmon mode resulting from its coupling to the normal component. We find that potential scattering mechanism is ineffective in wires with sufficiently large number of transverse channels (i.e., whose radii significantly exceed the interelectron distance), which corresponds to practically all up-to-date QPS experiments. The dissipation based mechanism leads to much higher QPS probability and, may possibly be an explanation of experimental observation of QPS events at sufficiently high values of bias currents \cite{beznat}.

Dynamics of low energy excitations of a superconducting wire is described by the following Lagrangian density
\begin{equation}\label{lag}
{\cal L}_0 = {\hbar\over 2}(i\rho\partial_\tau\phi+{I\over e}\partial_x\phi) + {\hbar^2C\over 8e^2}\left[(\partial_\tau\phi)^2 + c_s^2(\partial_x\phi)^2\right],
\end{equation}
where $\phi(x, \tau)$ is the phase of the superconducting order parameter. The variables $x$ and $\tau$ are the coordinate along the wire and the
imaginary (Matsubara) time respectively. Effective Lagrangian in Eq. (\ref{lag}) can be derived from
microscopics, e.g. Refs. \cite{zaikin, me}, as an expansion in powers of gradients of $\phi$. The last two terms describe propagating plasma mode \cite{ms}; here $C$ is capacitance (per unit length) of the wire and $c_s$ is the (Mooij-Schon) phase velocity \cite{ms}. Note that the plasma mode in Eq.(\ref{lag}) is gapless, which is a property of 1D wires: In sufficiently thin wires the screening turns out to be effectively weak because the electric field is ``pushed out'' to the region outside the wire. Note that Eq.(\ref{lag}) is valid only at sufficiently large distances (and times), greater than some cut-off length $\xi$, of the order of the superconducting coherence length. The term $i\hbar\rho\partial_\tau\phi/2$ is the so-called Berry phase \cite{com1}. The importance of this term has been discussed in Ref. \cite{ prokof, khleb2, khleb, me}: It accounts for momentum conservation during the phase slip formation, which leads to the suppression of the QPS rates (see below). The term proportional to $\partial_x\phi$ is due to the applied bias current $I$.

We start by considering a translationally invariant system. The rate of the QPS events can be evaluated by using the instanton method \cite{coleman}. We are interested in the decay probability of a current-carrying state $|I\rangle$, i.e.,  with bias supercurrent $I$. A corresponding instanton solution, i.e., a classical trajectory for the Lagrangian in Eq. (\ref{lag}) satisfying appropriate boundary conditions \cite{prokof, zaikin}, consists of a kink-antikink pair,
\begin{equation}\label{phi}
\phi_c(x, \tau) = \tan^{-1}{x-x_1\over c_s(\tau-\tau_1)}-\tan^{-1}{x-x_2\over c_s(\tau-\tau_2)}\ ,
\end{equation}
and the rate is given by the expression
\begin{equation}\label{rate0}
\Gamma_{\rm QPS}= K \,{\rm Im}\,\int d(\tau_1-\tau_2)\int dx_1 dx_2 e^{-{\cal S}(\phi_c)/\hbar} \ .
\end{equation}
The prefactor  $K$ in Eq. (\ref{rate0})  is related to the so-called fluctuation determinant \cite{coleman}, whose value can only be estimated, e.g. Ref. \cite{khleb2}, $K\sim c_s^2/\xi^4$. The classical action, corresponding to the classical (instanton) trajectory from Eq. (\ref{phi}), is readily obtained as ${\cal S}(\phi_c)=\int d\tau dx\, {\cal L}_0 (\phi_c)$,
\begin{equation}\label{act}
{{\cal S}(\phi_c)\over \hbar} = \alpha\ln{{\Delta x^2+c_s^2\Delta\tau^2}\over \xi^2} + {\pi I\over e}\Delta\tau + i\pi \rho \Delta x.
\end{equation}
In Eq. (\ref{act}) $\alpha = \pi\hbar Cc_s/4e^2$ and $\Delta x= x_1-x_2$, $\Delta \tau=\tau_1-\tau_2$. It is convenient to carry out the $\Delta\tau$ integration in Eq. (\ref{phi}) and then do the
remaining $\Delta x$ integral. The $\Delta\tau$ integral is formally divergent and one first needs to analytically continue it by bending the contour of integration into the complex plane in the region of negative argument \cite{coleman}. As a result one obtains
\begin{eqnarray}\label{int}
{\rm Im}\, \int  {d\Delta\tau e^{\pi I \Delta\tau}\over (\Delta x^2+c_s^2\Delta\tau^2)^\alpha} = {\pi^{3/2}\over c_s \Gamma(\alpha)} \left({\pi I\over 2ec_s \Delta x}\right)^{\alpha-1/2}\\\nonumber
\times J_{\alpha-1/2}(\pi I\Delta x/ec_s).
\end{eqnarray}
The subsequent integral over $\Delta x$ yields $0$. To see this one may use identity \cite{grad}
\begin{equation}\label{ident}
J_\nu(z)= {(z/2)^\nu\over \Gamma(\nu+1/2)\Gamma(1/2)}\int_0^\pi d\theta \sin^{2\nu}{\theta}\cos{(z\cos{\theta})}.
\end{equation}
Thus, replacing the Bessel function in Eqs. (\ref{int}) by its integral representation, Eq. (\ref{ident}), the integral over $\Delta x$ in Eq. (\ref{rate0}) yields $\delta (ec_s\rho \pm I \cos{\theta})$, and
therefore the remaining integral over $\theta$ is nonzero only for $I\ge ec_s\rho$. It is easy to verify that for a superconducting wire quantity $ec_s\rho$ always greatly exceeds the value of critical current \cite{beznat} and therefore $\Gamma_{\rm QPS}=0$. That is, the momentum released as a result of a phase slip event needs to be absorbed by the plasmons. Such process, however, is suppressed by the Landau criterion \cite{khleb}, i.e., impossibility to transfer both momentum and energy from a superflow to a bath of excitations (i.e., the plasmons), whose spectrum is linear, e.g. Eq.(\ref{lag}).

If translational invariance of the system is broken, the momentum conservation prohibiting QPS formation is ``violated'' and, as a result, the QPS rate becomes nonzero \cite{prokof, khleb}. The presence of lattice or disorder (in the following we will study the effects of the latter) obviously modifies the Lagrangian density in Eq. (\ref{lag}).  In the second order terms the disorder leads to the renormalization of the stiffness of the plasmon mode, e.g. Ref. \cite{zaikin}. Such renormalization, i.e., averaging over the disorder realizations, is well justified since the disorder potential varies on a scale small compared to cut-off length $\xi$. The contribution from the Berry phase term, however, comes from the discontinuity of the phase at the discrete set of points, e.g., QPS centers, as we have seen above. Therefore the electron density $\rho$ in this term does not self-average, i.e., one should explicitly account for its spacial variation. Since the classical equation of motion is independent of the Berry phase term, the instanton solution and the rate are still given by  Eqs. (\ref{phi}, \ref{rate0}), while the classical action becomes
\begin{equation}\label{act1}
{{\cal S}(\phi_c)\over \hbar} = \alpha\ln{{\Delta x^2+c_s^2\Delta\tau^2}\over \xi^2} + {\pi I\over e}\Delta\tau + i\pi \int_{x_1}^{x_2}dx \rho(x). \end{equation}
The only distinction between Eq. (\ref{act1}) and  Eq. (\ref{act}) is obviously the last term. By writing $\rho(x)=\rho_0+\delta\rho(x)$ and assuming that $\rho_0\gg\delta\rho$ (which is a well justified assumption for a wire whose diameter is much greater than the interelectronic distance, - recall that $\rho$ is 1D density, i.e., averaged over the wire's cross-section), we rewrite the $x$ integration in Eq. (\ref{rate0}) as
\begin{eqnarray}\label{int2}
\int dx_1 dx_2 e^{i\pi\int_{x_1}^{x_2}dx \rho(x)} F(x_1-x_2) \simeq  (L /\rho_0^2)
\\\nonumber
\times\int d\Delta x \langle\delta\rho(\Delta x)\delta\rho(0)\rangle F(\Delta x)e^{i\pi\rho_0\Delta x} ,
\end{eqnarray}
where $F$ is the remaining $x$-dependent part of the integrand function in Eq. (\ref{rate0}). In deriving Eq. (\ref{int2}) (together with the inequality $\rho_0\gg\delta\rho$) we used a fact that $F(\Delta x)$ varies in a scale $c_s/I$, which is much greater than $\rho_0^{-1}$.

Then, substituting Eqs. (\ref{int}, \ref{int2}) into Eq. (\ref{rate0}) and using Eq. (\ref{ident}), after some straightforward algebra, one obtains
\begin{equation}\label{rate1}
\Gamma_{\rm QPS} = {\pi K \xi^2 L\over c_s \Gamma(2\alpha)}\times\left({\pi I \xi\over ec_s}\right)^{2\alpha-1}\times{\langle|\delta\rho_{{\bf p}_0}|^2\rangle\over \xi\rho_0^2},
\end{equation}
where $\langle|\delta\rho_{{\bf p}}|^2\rangle =\int dx \langle\delta\rho(x)\delta\rho(0)\rangle e^{ipx}$ and $p_0=\pi\rho_0$.

The first two factors in the rhs of Eq. (\ref{rate1}) can be put in a traditional form $\omega_0e^{-S_0}$, where $\omega_0\sim c_s L/\xi^2$ is an attempt frequency and $S_0=(2\alpha-1)\ln{(ec_s/\pi I \xi)}$. This result has essentially been obtained in Ref. \cite{zaikin}. The third factor (which we will denote $P_3$ below) is a correction due to the account of the Berry phase. It can be viewed as a probability of the condensate to change its momentum by $2\pi (\rho_0/2)$ due to an external potential. Indeed, by using a linear response relation $\delta\rho^{3D}_{\bf p} \sim \chi_{\bf p} V_{\bf p}$ \cite{com}, where $\chi_{\bf p}$ is the static susceptibility of the electron gas,  $\rho^{3D}_{\bf p}$ is conventional (3D) conduction electron density and $V_{\bf p}$ is the disorder potential, one has $\langle|\delta\rho_{{\bf p}_0}|^2\rangle = A|\chi_{{\bf p}_0}|^2 |V_{{\bf p}_0}|^2$, where $A$ is the cross-section area of the wire. (Thus $P_3$ is proportional to the Born's scattering amplitude.)

Since momentum $p_0$ is very high compared to both the Fermi momentum and the inverse interatomic distance (recall that $p_0=\pi\rho^{3D}A$), we can expect that $P_3$ is rather small. In order to estimate it, one needs to specify $V_{\bf p}$ and $\chi_{\bf p}$. At sufficiently high momenta the latter can be approximated by the susceptibility of a free electron gas,
$\chi_{\bf p}\simeq\chi^{\rm free}_{{\bf p}\rightarrow\infty}\simeq 2k_F^5/(3\pi^2 E_F p^2)$, where $k_F$ and $E_F$ are Fermi momentum and energy ($E_F=\hbar^2k_F^2/2m$). For the disorder potential we assume that $\langle V({\bf r})V({\bf r}')\rangle=V_0^2e^{-|{\bf r}-{\bf r}'|/a}$, where the value of $a$ is of the order of interatomic distance, a few $\AA$. The strength of the disorder $V_0$ can be estimated from the mean free path formula, $v_F/l=(\chi^{\rm free}_{{\bf p}=0}/4)\int d\Omega (1-\cos\theta)\langle|V_{{\bf p}-{\bf p}^\prime}|^2\rangle$, where $\theta$ is angle between momenta ${\bf p}$ and ${\bf p}^\prime$, whose magnitudes are equal to $k_F$, and $\Omega$ is solid angle. Evaluating the Fourier transform of $\langle V({\bf r})V({\bf r}')\rangle$ we have $\langle|V_{{\bf p}}|^2\rangle = 8\pi a^3V_0^2/(1+a^2 p^2)^2$, and, from the above mean free path formula we obtain that $V_0^2/a \simeq (\hbar k_F)^4/[2m^2 l \ln{(2ak_F)}]$, where we assumed that $(2ak_F)^2\gg 1$. Then, we have
\begin{equation}\label{p3}
P_3\simeq {64\over 9\pi}{A\over \xi l \ln{(2ak_F)}}\left({k_F\over p_0}\right)^{10}.
\end{equation}
Note that $p_0/k_F$ is, up to a numerical factor, the number of transverse Fermi channels ($N_\perp$) for the wire.

It is instructive to estimate the QPS rate for a typical experimental system, such as $10\, nm$ thick and $200 \, nm$ long MoGe nanowire. The value of conduction electron density $\rho^{3D}=k_F^{3}/3\pi^2$ can be obtained from the data on conductivity in the normal state. Using  Drude formula $\sigma = e^2\rho^{3D}\tau_{\rm el}/m$, $\tau_{\rm el} = l/v_F$, where the mean free path $l$ is assumed to be of the order of interatomic distance, $l\simeq 4\AA$ \cite{beasley}, for $\sigma^{-1} \simeq 2\, \mu\Omega\, m$ \cite{zaikin2}, we find $k_F\simeq 1\,\AA^{-1}$ and $\rho^{3D}\simeq 3\times 10^{28}\, m^{-1}$. Then, for $\xi=8 \,nm$ \cite{bez} and $a=4 \AA$ we obtain that $P_3\sim 10^{-27}$.
The phase velocity $c_s$, which enters the first two factors in Eq. (\ref{rate1}), can be expressed in terms of the penetration depth $\lambda$ as $c_s=c r_0/\lambda (C)^{1/2}$ \cite{zaikin, ms}, where $r_0= 5 \, nm$ is the cross section radius, $c$ is the speed of light and the capacitance $C$ is  given by $C\simeq 2\pi\epsilon [\ln{(L/r_0)}]^{-1}$, ($\epsilon$ is the dielectric constant of surrounding medium). For a typical experimental situation $\epsilon\sim 5$ \cite{rog} and so $C\simeq 9$. For MoGe $\lambda\simeq 0.72\,\mu m $ \cite{zaikin2}, which gives $c_s\simeq 7\times 10^5\, m/s$ and $\alpha\simeq 2.4 $. Then we find that the attempt frequency, i.e., the first factor in Eq.(\ref{rate1}) is $\omega_0\sim 10^{15} - 10^{16}\, m/s$. The value of $S_0$ decreases with the growth of the bias current. For $I\sim 2\mu A$, which is of the order of critical current in a $10\, nm$ thick MoGe wire \cite{beznat} we obtain $S_0\simeq 3.2$. Then, according to Eq.(\ref{rate1}) $\Gamma_{QPS}$ does not exceed $10^{-12} s^{-1}$. Thus we conclude that the mechanism of condensate scattering at the disorder potential, e.g. Refs. \cite{prokof, khleb, khleb2} is ineffective even in such disordered systems as MoGe nanowires.

Note that the above estimate breaks down for the wires with the number of transverse channels $N_\perp$ of order 1. (For the above parameters $N_\perp \sim 100$.) Also, in the presence a weak link (a region with $N_\perp \leq 1$ due to the inhomogeneity in wire's cross section, etc.) the main contribution into the integral in the QPS rate in Eq.(\ref{rate0}) comes from the vicinity of point $x_{\rm link}$ with $\rho(x_{\rm link})\simeq 0$. In that case one may set $\Delta x \simeq 0$ ($x_1,\, x_2\simeq x_{\rm link}$) in Eq. (\ref{act1}) and the QPS rate corresponds to that for a Josephson junction with an effective fugacity of the kinks $\sim |\int dx_1 \exp{[i\pi \int_{-\infty}^{x_1}dx \rho(x)]}|$.

In the remaining part of the paper we investigate another source for the QPS production based on dissipation of the plasmon mode due to the normal component of the electronic liquid. Such mechanism is motivated by the report of experimental observation of the QPS at high bias currents, i.e., when the value of the bias current is close to critical current \cite{beznat}. At such high currents superconductors can become gapless \cite{tinkam}, which leads to the appearance of appreciable normal component even at temperatures much lower than $T_c$. In order to estimate the effect of dissipation we recall that a superconducting wire can be described by an effective transmission line, which one may represent
as a ladder of elementary building blocks with inductors (or Josephson junctions) along one stringboard of the ladder and capacitors on the rungs. It is easy to show that in the continuous limit such model reproduces wave equation corresponding to the classical equation of motion for the Lagrangian in Eq. (\ref{lag}). The presence of the normal component can be modeled by adding resistors connected in parallel to the inductors. A straightforward analysis of the circuit shows that  such resistors introduce dissipation according to the relation $dE/dt = \hbar^2 /(8e^2R_n)\int dx(\partial^2\phi/\partial\tau\partial x)^2$, where $R_n$ is effective resistance (per unit length) of the circuit and we have used the familiar relation between the voltage and the phase, $\Delta V =\hbar \Delta\phi/2 e$. Such dissipative function corresponds to an additional term in the action of the wire, which in the imaginary time representation can be cast in the form
\begin{equation}\label{diss}
{{\cal S}_{\rm diss}\over \hbar} = {\hbar\sigma_n A\over 8 e^2}\int {d\omega\over 2\pi}\, |\omega| |\partial_x\phi(x,\omega)|^2,
\end{equation}
where $\sigma_n = 1/R_n A$ is the conductivity of the normal component, $\sigma_n= e^2\rho^{3D}_n\tau_{\rm el}/m$ and $\rho^{3D}_n$ is density of normal electrons. Together with ${\cal L}_0 $ this term yields familiar dispersion relation for the plasmon mode, $\omega^2 - c_s^2k^2+i\omega c_s^2 k^2/\omega_d=0$, e.g. Ref. \cite{ms}.

The QPS rate in the presence of dissipation can be evaluated perturbatively in ${\cal S}_{\rm diss}$. The first order term is
\begin{equation}\label{rate2}
\Gamma_{\rm QPS}= K \,{\rm Im}\,\int d\Delta\tau\int dx_1 dx_2 e^{-{\cal S}(\phi_c)/\hbar}\,{\cal S}_{\rm diss}(\phi_c)/\hbar \, ,
\end{equation}
where $\phi_c$  and ${\cal S}(\phi_c)$ are given by Eqs. (\ref{phi}, \ref{act}), - here we neglect neglect the spatial variation of $\rho$, e.g., Eq. (\ref{act1}). After a straightforward evaluation of integral in Eq. (\ref{diss}) we obtain that
\begin{equation}\label{actdiss}
{\cal S}_{\rm diss}(\phi_c) = {\pi\hbar^2\rho^{3D}_n A\tau_{\rm el}\over 2m}{|\Delta x|^3\over (\Delta x^2+c_s^2\Delta\tau^2)^2}.
\end{equation}
Then, integration over $\Delta\tau$ can again be performed with the use of Eqs. (\ref{int}) with an obvious replacement $\alpha\rightarrow\alpha -2$,
\begin{eqnarray}\label{rate21}
\nonumber
\Gamma_{\rm QPS}= {\pi^{3/2} K\xi^2 L \over 4c_s\Gamma (\alpha+2)}\times \left({\pi I \xi\over ec_s}\right)^{2\alpha-1}\times
{\pi\hbar\rho^{3D}_n A\tau_{\rm el}\over 2m\xi}\,
\\
\int_0^\infty dz
{J_{\alpha+3/2}\over z^{\alpha - 3/2}}\cos{\left({e\rho c_s\over I}z\right)}  .
\end{eqnarray}
The first two factors in Eq. (\ref{rate21}) are identical (up to a numerical factor) to those in Eq. (\ref{rate1}). The integral in the third factor can be evaluated by applying identity (\ref{ident}). Note that unlike the previous case, e.g. Eqs. (\ref{rate0} - \ref{int}), the integrand function now contains an additional factor $z^3$. Because the power of $z$ is odd (thanks to the odd power of $|\omega|$ in ${\cal S}_{\rm diss}$ in Eq.(\ref{diss})), the $z$-integration no longer produces $\delta (ec_s\rho/I\pm \cos{\theta})$ (or its derivatives), which was the case in Eq. (\ref{rate0}) and, as a result, led to the zero rate for $c_s\rho > I/e$. Evaluating the integral in the limit $c_s\rho \gg I/e$ we obtain that the third factor ($P_3^\prime$) in Eq. (\ref{rate21})
\begin{equation}\label{p31}
P_3^\prime\sim{\hbar\rho^{3D}_n A\tau_{\rm el}\over m\xi}\, \left({I\over ec_s\rho}\right)^4.
\end{equation}
The rapid oscillations of the integrand function in Eq.(\ref{rate21}) again lead to significant suppression of the rate. An estimate gives $P_3^\prime\simeq 10^{-19}$, where we have assumed again that $I=2\mu A$, which is a typical value in experiments in Ref. \cite{beznat}) and that $\rho^{3D}_n\sim \rho^{3D}$. This gives $\Gamma_{\rm QPS}\sim 10^{-4}\, s^{-1}$. While this estimate is 4 orders of magnitude lower than the experimentally observed QPS rate \cite{beznat}, it is $8$ orders of magnitude greater than the previous one, e.g. Eqs.(\ref{rate1}, \ref{p3}). Moreover, in Eq.(\ref{rate21}) we did not account for the softening of the plasmon mode. Indeed, since $c_s\sim (\rho_s)^{1/2}$ \cite{zaikin, ms}, where $\rho_s$ is the density of superconducting electrons, $c_s$ is expected to drop to $0$ at the transition point (i. e., where $\rho_n=\rho$ and $\rho_s=0$), which may lead to substantial increase in $P_3^\prime$, etc. The discrepancy may also be related to other, more efficient sources of dissipation, such as shunting or contact resistances. A proper account of finite size of the nanowires may also lead to significant modifications of the QPS rates on both qualitative and quantitative levels \cite{buch}.

Finally we emphasize that the dissipation based mechanism relies on the assumption that the fraction of the normal component is of order 1. Such an assumption obviously breaks down for a conventional (BCS) superconductor, where at sufficiently low temperatures the density of the normal component is exponentially suppressed by the gap in the excitation spectrum. The normal component (and thus the relaxation rate of the plasmon mode) can be significantly enhanced by pair-breaking mechanisms, such as high bias current (as in Ref. \cite{beznat}). Note that the plasmon resonances in superconducting nanowires can be studied experimentally by measuring the reflectivity coefficients of the wires, e.g., Ref. \cite{plasma}. Such measurements could therefore test the validity of arguments presented in this Letter.

In summary we have shown that a proper account of the momentum conservation during the QPS formation leads to a very strong suppression of
the QPS rate, e.g. by factor $P_3$ in Eqs.(\ref{rate1}, \ref{p3}) (for the potential scattering mechanism) and $P_3^\prime$ in Eqs.(\ref{rate21}, \ref{p31}) (for the dissipation based mechanism). The latter mechanism, under special circumstances, may lead to an observable QPS rate and may be the origin of the enhancement of phase slip transitions observed in Ref. \cite{beznat}.

{\acknowledgements The author thanks A. Rogachev for valuable discussions. The work is supported by the US DOE.}


\end{document}